\documentclass[11pt]{article}
\usepackage{amsmath,graphicx}

0.1cm \evensidemargin - 5cm

\begin{document}

\title{{\Large {\textbf{Phase diagram of a 2D Ising model within a nonextensive approach}}}}
\author{{D. O. Soares-Pinto}{$^{1,2}$}\thanks{\emph{Electronic
address:} \texttt{dosp@cbpf.br}} , {I. S. Oliveira}{$^{1}$} and
{M. S. Reis}{$^{2}$}}
\date{}
\maketitle

\begin{center}
$^{1}$ \textit{Centro Brasileiro de Pesquisas F\'{\i}sicas, Rua Dr. Xavier Sigaud 150, 22290-180, Rio de Janeiro, Brazil}%
\\[0pt]
$^{2}$ \textit{CICECO, Universidade de Aveiro, 3810-193, Aveiro, Portugal}\\[0pt]
\end{center}

\begin{abstract}
In this work we report Monte Carlo simulations of a 2D Ising
model, in which the statistics of the Metropolis algorithm is
replaced by the nonextensive one. We compute the magnetization and
show that phase transitions are present for $q\neq 1$. A
$q\,-$\,phase diagram (critical temperature vs. the entropic
parameter $q$) is built and exhibits some interesting features,
such as phases which are governed by the value of the entropic
index $q$. It is shown that such phases favors some energy levels
of magnetization states. It is also showed that the contribution
of the Tsallis cutoff is essential to the existence of phase
transitions.
\end{abstract}
%
%


\tableofcontents

\newpage

\section{Introduction}

The nonextensive statistics is a generalization of the Boltzmann
-- Gibbs one and it is based on the nonadditive entropy
\cite{1988_JSP_52_479}

\begin{equation}\label{eq.01}
S_{q} = k \frac{1-\sum_{i} p_{i}^{q}}{q-1}\qquad (q\,\in\, \Re)
\end{equation}
where $q$ is the entropic index for a specific system, connected
to its dynamics, as recently proposed
\cite{2006_PRB_73_092401,2006_EPJB_50_99}; $p_{i}$ are
probabilities satisfying $\sum_{i} p_{i}=1$, $k$ is a constant,
and $\lim_{q\rightarrow 1}S_{q} = S_{BG}$, where $S_{BG}$ is the
Boltzmann -- Gibbs entropy. In this statistics, a system composed
of two independent parts $A$ and $B$, in the sense that the
probabilities of the systems factorize, has the following
pseudo-additivity (nonextensivity) property of the entropy
\cite{1999_BrJP_29_1,2004_CMT_16_223}

\begin{equation}\label{eq.02}
S_{q}(A+B) = S_{q}(A)+S_{q}(B)+(1-q)\,S_{q}(A)\,S_{q}(B)/k.
\end{equation}
This pseudo-additivity is related to the composability property of
$S_{q}$ \cite{livro_TsallisCap1}. Since for any system $S_{q}\geq
0$, then $q<1$ correspond to superadditivity (superextensivity),
$q=1$ to additivity (extensivity), and $q>1$ to subadditivity
(subextensivity). Besides representing a generalization, $S_{q}$,
as much as $S_{BG}$, is positive, concave and Lesche-stable ($\forall\,
q > 0$). Recently, it has been shown that it is also extensive for
some kinds of correlated systems in which scale invariance
prevails \cite{2005_PNAS_102_15377,website_TEMUCO}.

In this paper we report some results of a Monte Carlo simulation
of a 2D Ising model upon replacing the statistics of the
Metropolis algorithm by the nonextensive statistics. From
numerical calculations we compute the magnetization of the system,
and built a $q\,-$\,phase diagram showing that, even for $q\neq 1$,
exist phase transitions, in contrast to a previous work
\cite{1999_PhysA_268_553}. The text is organized as follows: In
Section 2, we describe the equilibrium distribution of
nonextensive statistics and the importance of the internal energy
constrains. In Section 3, we describe the introduction of the
nonextensive formalism into the Monte Carlo method. In Sections 4
and 5, we discuss the main results and describe the behavior of
the critical temperature with the entropic index in a phase
diagram ($T_{c}$ vs. $q$).

\section{Nonextensive statistics}

To calculate the equilibrium distribution, the above entropy,
Eq.(\ref{eq.01}), must be maximized \cite{1988_JSP_52_479}. If the
system is isolated, i.e., in a microcanonical ensemble,

\begin{equation}\label{eq.03}
\sum_{i=1}^{\Omega}p_{i} = 1
\end{equation}
the maximization yields equiprobability of states occupation. On
the other hand, if the system is in contact with a thermal
reservoir (canonical ensemble), it is necessary to add the
internal energy constraints, which can be done according to three
possible choices. The first one is \cite{1988_JSP_52_479}

\begin{equation}\label{eq.04}
\sum_{i=1}^{\Omega}p_{i}\,\varepsilon_{i} = U
\end{equation}
the standard definition of internal energy in which
$\{\varepsilon_{i}\}$ are the eigenvalues of the Hamiltonian of
the system. The second, as postulated in \cite{1991_JPA_24_69}, is:

\begin{equation}\label{eq.05}
\sum_{i=1}^{\Omega}p_{i}^{q}\varepsilon_{i} = U_{q}.
\end{equation}
Both definitions presents some difficulties with the interpretation
of some results \cite{1998_PhysA_102_15377}. Thus, a third choice for
the internal energy constraint was introduced as
\cite{1998_PhysA_102_15377}:

\begin{equation}\label{eq.06}
\frac{\sum_{i=1}^{\Omega}p_{i}^{q}\varepsilon_{i}}{\sum_{i=1}^{\Omega}p_{i}^{q}}
= U_{q}
\end{equation}
which defined the escort probability (first introduced in Ref.
\cite{livro_beck}):

\begin{equation}\label{eq.07}
P_{i}^{(q)} \equiv \frac{p_{i}^{q}}{\sum_{j=1}^{\Omega}p_{j}^{q}}.
\end{equation}
The maximization of $S_{q}$ in that case yields for the probability
distribution

\begin{equation}\label{eq.08}
p_{i} = \frac{1}{Z_{q}}\,
\left[1-\frac{(1-q)\,\beta\,(\varepsilon_{i}-U_{q})}{\sum_{j=1}^{\Omega}p_{j}^{q}}\right]^{1/(1-q)}
\end{equation}
where

\begin{equation}\label{eq.09}
Z_{q} =
\sum_{i=1}^{\Omega}\left[1-\frac{(1-q)\,\beta\,(\varepsilon_{i}-U_{q})}{\sum_{j=1}^{\Omega}p_{j}^{q}}\right]^{1/(1-q)}
\end{equation}
is the nonextensive partition function and $\beta$ a Lagrange
multiplier. After some algebraic manipulations
\cite{1998_PhysA_102_15377} it becomes:

\begin{equation}\label{eq.10}
p_{i} = \frac{1}{Z_{q}^{\prime}}\,
\left[1-(1-q)\,\beta^{\prime}\,\varepsilon_{i}\right]^{1/(1-q)} =
\frac{1}{Z_{q}^{\prime}}\,e_{q}^{-\beta^{\prime}\varepsilon_{i}}
\end{equation}
and

\begin{equation}\label{eq.11}
Z_{q}^{\prime} =
\sum_{j=1}^{\Omega}\left[1-(1-q)\,\beta^{\prime}\,\varepsilon_{j}\right]^{1/(1-q)}
= \sum_{j=1}^{\Omega}e_{q}^{-\beta^{\prime}\varepsilon_{i}}
\end{equation}
where

\begin{equation}\label{eq.12}
\beta^{\prime} =
\frac{\beta}{\sum_{j=1}^{\Omega}p_{j}^{q}+(1-q)\,\beta\,U_{q}}.
\end{equation}
and $e_{q}^{x}$ is the generalized exponential, which has the
following the property:

\begin{equation}\label{eq.13}
[1-(1-q)\,\beta^{\prime}\varepsilon_{i}]^{1/(1-q)} = \left\{
\begin{array}{ll}
         e_{q}^{-\,x}, & \mbox{if $1-(1-q)\,x \geq 0$};\\
         0, & \mbox{if $1-(1-q)\,x < 0$}.\end{array} \right.
\end{equation}
known as the Tsallis cutoff procedure. A detailed
discussion about the role of constraints within the nonextensive
statistics was done by Tsallis et al \cite{1998_PhysA_102_15377},
but recently it has been showed by Ferri et al \cite{2005_JSMTE_04_009}
the equivalence of all these formulations of internal energy
constraints. In spite of that, in this work, to avoid
misunderstanding, we choose the normalized internal
energy form.

The thermal equilibrium in the nonextensive statistics is still an
open issue due to the definition of the physical temperature
\cite{2000_PhysA_283_59,2001_PhysA_295_246,2001_PhysA_295_416,2002_PhysA_305_52,2003_PhysA_317_209,2004_EPL_65_606,2006_PhysA_368_430}.
Thus, differently from some authors \cite{1999_PhysA_268_553}, in
our approach the parameter $\beta^{\prime}$ is assumed to be the
physical temperature, i.e., $\beta^{\prime} = (k\,T)^{-1}$. The
validity of this choice was first shown experimentally
\cite{2002_EPL_58_42}, and latter theoretically
\cite{2006_PRB_73_092401,2006_EPJB_50_99,2002_PRB_66_134417,2003_PRB_68_014404}
for manganites.

\section{Monte Carlo simulations of a 2D Ising model using nonextensive statistics}

In this section we are going to discuss the modification of the
Metropolis method for the nonextensive statistics, considering a
ferromagnetic 2D Ising with first-neighbors interaction. The
Hamiltonian is given by

\begin{equation}\label{eq.14}
\mathcal{H} = -J\sum_{\langle ij\rangle}s_{i}s_{j}
\end{equation}
where $\langle ij\rangle$ denotes the sum over first neighbors on a
square lattice of size $N = L\times L$, $s_{i} = \pm 1$ and $J > 0$
(ferromagnetic interaction). We proceed the single flip Monte Carlo
calculations \cite{livro_compphys} to obtain the magnetization of
the system, however we have changed the usual statistical weight to:

\begin{equation}\label{eq.15}
w = \frac{P_{i,after}^{(q)}}{P_{i,before}^{(q)}}=
\left[\frac{e_{q}^{-\varepsilon_{i}^{after}/k\,T}}
{e_{q}^{-\varepsilon_{i}^{before}/k\,T}}\right]^{q}
\end{equation}
or, in other words, the ratio between the escort probabilities
before and after the spin flip. Since this quantity is a ratio,
the normalization factor of the escort probabilities, i.e., the
generalized partition function, Eq.(\ref{eq.11}), cancels and the
weight calculated can be written as the ratio between the
generalized exponentials raised to the entropic parameter $q$. It
is important to emphasize that $w$  is the quantity that will be
compared to a random number in the Metropolis algorithm (see
appendix for details on the MC procedure used). It is also
important to note that the Tsallis cutoff procedure,
Eq.(\ref{eq.13}), must be taken into account, i.e., it must be
included into $w$ to avoid complex probabilities. The simulations
were done with the entropic parameter $q \in [0,1]$. Lattice
size were $L = 8, 16, 24, 32$, with periodical boundary
condition imposing that $s_{L+1}\equiv s_{1}$.

\section{Results and Discussion}

\begin{figure}[tbh]
\begin{center}
\includegraphics[width=0.4\columnwidth,angle=0]{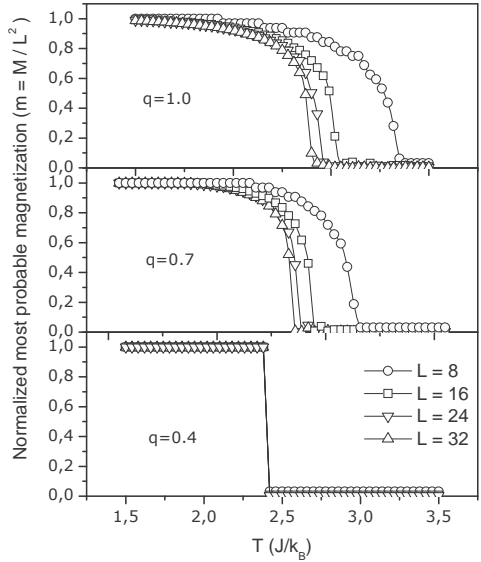}
\end{center}
\caption{Normalized most probable magnetization vs. temperature.
For $q=0.4$ it can be seen that the magnetization has no
dependence on the lattice size, dropping form $m=1$ to $m=0$
suddenly at $T_c$. It happens due to the contribution of the
Tsallis cutoff. However, for $q=0.7$ the magnetization depends on
the lattice size and smoothly goes to zero close to $T_c$ as a
second order phase transition. For $q=1$ it recovers the well
known result.}\label{Figure1}
\end{figure}

The most probable normalized magnetization, $m=M/L^{2}$, was
obtained after $5\times 10^{5}$ Monte Carlo steps and are shown on
Fig.(\ref{Figure1}) for $q = 0.4$, $q = 0.7$ (which are
representative results for $q < 0.5$ and $q > 0.5$, respectively),
and for $q = 1$. One can observe that for $q>0.5$ there are strong
influences of the lattice size on the shape of the magnetization
curve and on the critical temperature, $T_c$. In addition, the
magnetization drops smoothly to zero close to $T_c$ due to the
thermal fluctuations. On the contrary, for $q<0.5$, there are no
dependence of those quantities on the lattice size, and the
magnetization changes suddenly at $T_c$, from $m = 1$ to $m = 0$. In
other words, there are no thermal fluctuations in this case and the
magnetization works like a microcanonic two-level system.

This behavior, for $q<0.5$, is simple to be understood. At low
temperatures (for instance $T = 1.5$ and $q = 0.4$), the first
Monte Carlo steps lead the magnetization to $m = 1$, i.e., to the
ground state, as expected (due to the low temperatures). Then the
subsequent Monte Carlo steps attempt  to invert the spin, but it
fails because is energetically unfavorable. Then the Metropolis
algorithm takes place; as $\varepsilon_{i}^{after}=4\,J$ and
$\varepsilon_{i}^{before}=-4\,J$, therefore
$1-(1-q)\,\beta^{\prime}\,\varepsilon_{i}^{after}<0$ for all
$T<4\,J\,(1-q)$. So, considering the cutoff, Eq.(\ref{eq.13}),
$p_{i}^{after}=0$ and then $w=0$. Since the Metropolis algorithm
flips energetically unfavorable spins if the random number is
smaller then $w$, for $T<4\,J\,(1-q)$ those spins never flips ($w
= 0$), keeping the magnetization at $m = 1$; in other words, in
the ground state. This situation persists up to $T_{c} =
4\,J\,(1-q)$, where the cutoff for $p_{i}^{after}$ is no longer
satisfied and the thermal fluctuation can therefore acts. However,
the spins are already quite warm and the magnetization drops
suddenly to zero, i.e., to a equiprobable state. A similar
behavior was already found describing the generalized Brillouin
function \cite{2002_PRB_66_134417}.

\begin{figure}[tbh]
\begin{center}
\includegraphics[width=0.4\columnwidth,angle=0]{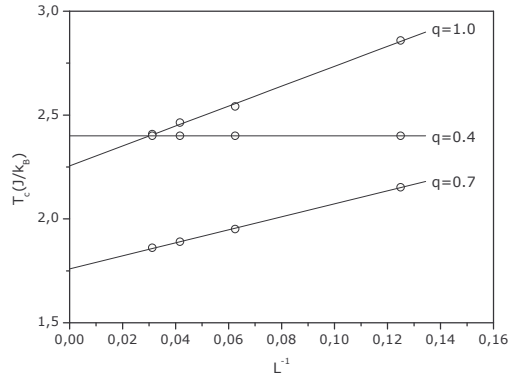}
\end{center}
\caption{Taking the thermodynamical limit of the system, one can
notice that the behavior is quite different for $q<0.5$ and for
$q>0.5$. As can be seen, for $q=0.4$ the critical temperature has
no dependence on the lattice size. It happens due to the
contribution of the Tsallis cutoff. However, for $q=0.7$ the
lattice size dependence appears. For $q=1$ the thermodynamical
limit recovers the critical temperature of Onsager result
($T_{c}=2\,[\arctan(1/\sqrt{2})]^{-1}$).}\label{Figura2}
\end{figure}

To determine the critical temperature we must take the
thermodynamic limit ($L\rightarrow\infty$). In Fig.(\ref{Figura2})
we plot the critical temperature as a function of the inverse
lattice size for different values of the entropic parameter,
taking the limit ($L^{-1}\rightarrow 0$). Notice that the critical
temperature for $q = 1$ tends to Onsager result and, as explained
above, it does not depend on the lattice size for $q < 0.5$.
Similar independency were found in different systems
\cite{2000_PhysA_286_156}. Also, the slope of the curves changes
with the entropic parameter $q$, suggesting a dependence of the
critical exponents on the entropic parameter (for studies about
this connection see for example
\cite{1999_PRL_83_4233,2000_EPJB_17_679,2001_PhysA_290_159}).

\begin{figure}[tbh]
\begin{center}
\includegraphics[width=0.4\columnwidth,angle=0]{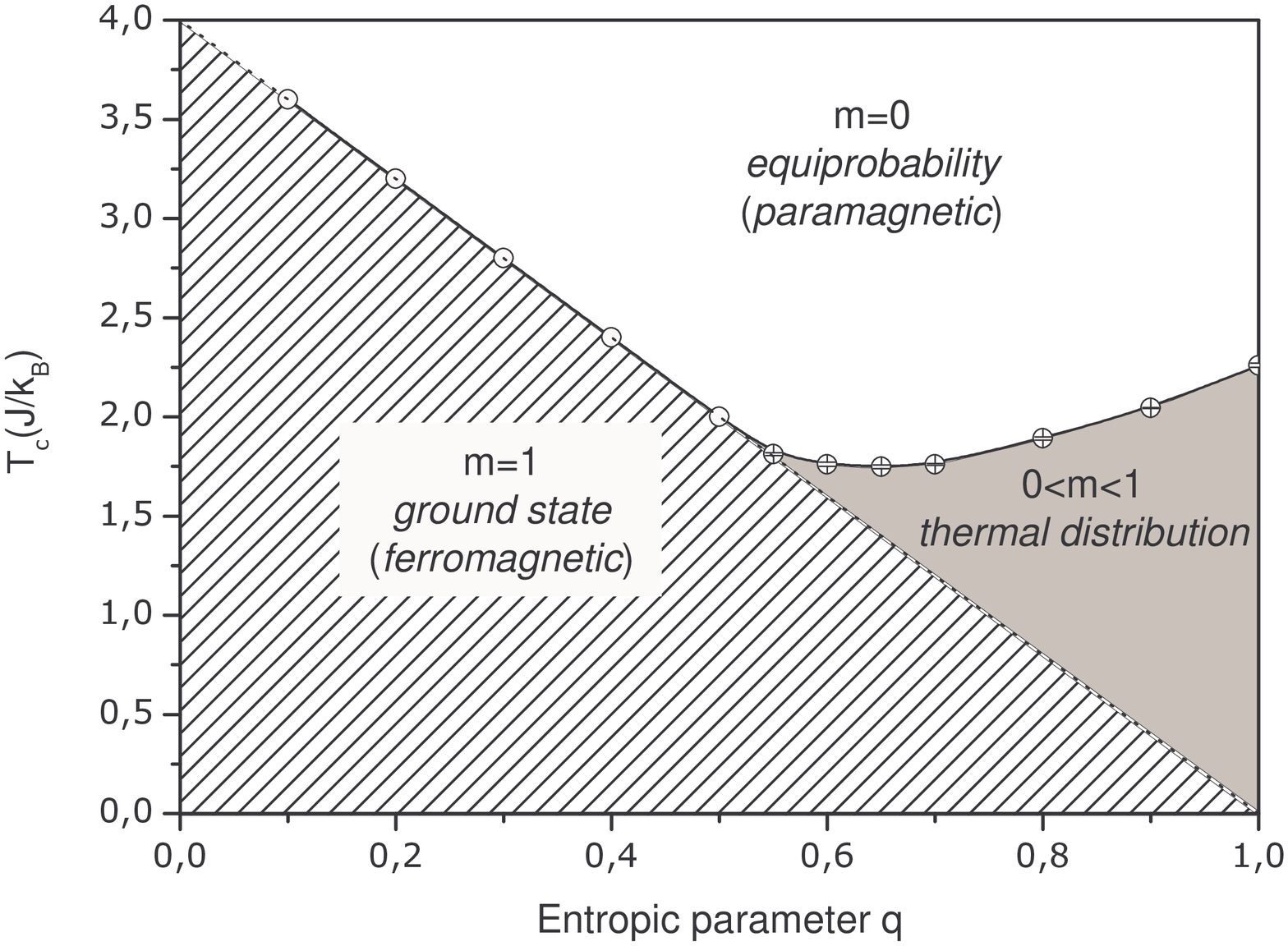}
\end{center}
\caption{Three different regions can be seen in this $q\,-$\,phase
diagram. The first one is the ferromagnetic ground state ($m=1$)
which is kept from zero Kelvin until $T_{c}=4\,J\,(1-q)$, due to
the contribution of the Tsallis cutoff. The second region has
influence at the thermal distribution ($0<m<1$). The third is the
equiprobable paramagnetic state ($m=0$). The error is smaller then
the the size of those dots.}\label{Figura3}
\end{figure}

With those critical temperature values we build a phase diagram
shown in Fig.(\ref{Figura3}), i.e., $T_{c}$ as a function of $q$. It
is quite interesting because, in contrast to previous works
\cite{1999_PhysA_268_553}, we found that for the Monte Carlo
simulations of a 2D Ising model in nonextensive statistics has phase
transitions for $q\neq 1$. It is clear that below the $4\,J\,(1-q)$
line the system is in the ground state and then $m = 1$. Above this
line there are two regions where the thermal fluctuation act: one
above $T_c$, i.e., in the paramagnetic regime and, consequently, in
the equiprobable state; and the other regime lies between
$4\,J\,(1-q)$ line and $T_c$ when the magnetization assume values
between 0 and 1. It is interesting that the slope of the critical
temperature, $T_{c}(q)$, does not changes abruptly with the increase
of the entropic parameter. For $q \sim 0.5$ the slope changes
smoothly indicating that the spin is not warm enough and pass to a
thermal distribution region before the equiprobable state.

\section{Conclusions}
In this work, we studied a ferromagnetic 2D Ising model with
first-neighbor interactions through a Monte Carlo simulation in
which the Metropolis algorithm was changed to the nonextensive
statistics. Magnetization as a function of the temperature for
different values of $q$ were evaluated and we found phase
transition for $q\neq 1$, in contrast to a previous work
\cite{1999_PhysA_268_553}. This results arises due to the
definition of the physical temperature. In addition, we also have
showed the contribution of the Tsallis cutoff is of great
importance and rules the phase transition for $q < 0.5$.

\bigskip

The authors acknowledge S.M.D. Queir{\'{o}}s, C. Tsallis and R.
Toral for their comments. We would like to thanks the Brazilian
funding agencies CNPq and CAPES. DOSP would like to thanks the
Brazilian funding agency CAPES for the financial support at
Universidade de Aveiro at Portugal.

\appendix
\section*{Appendix}
Each Monte Carlo step can be resumed as the following

\begin{enumerate}
\item Compute the interaction energy of a given spin $i-$th of the
lattice with its neighbors $\varepsilon_{i}^{before} =
\sum_{j=1}^{4}\varepsilon_{ij}$. After that, change the state of
this spin and compute again its interaction energy,
$\varepsilon_{i}^{after}$. If $\varepsilon_{i}^{after}<
\varepsilon_{i}^{before}$, accept the change of state;

\item If the energy is not lower, using Eq.(\ref{eq.13}), compute
$w$. Compare this quantity to a number that belongs to the
interval $[0,1]$ generated randomly. Being this random number
smaller then $w$ then accept the change of state, otherwise not.

\end{enumerate}
As can be seen, this is the ordinary Metropolis algorithm in which
the probability of state was changed from the Boltzmann weight to
the Tsallis factor, Eq.(\ref{eq.07}).


\begin{thebibliography}{10}

\bibitem{1988_JSP_52_479}
C.~{Tsallis}.
\newblock {Possible generalization of Boltzman-Gibbs statistics}.
\newblock {\em Journal of Statistical Physics}, 52:479, 1988.

\bibitem{2006_PRB_73_092401}
M.~S. {Reis}, V.~S. {Amaral}, R.~S. {Sarthour}, and I.~S.
{Oliveira}.
\newblock {Experimental determination of the nonextensive entropic parameter
  q}.
\newblock {\em Physical Review B}, 73:092401, 2006.

\bibitem{2006_EPJB_50_99}
M.~S. {Reis}, V.~S. {Amaral}, R.~S. {Sarthour}, and I.~S.
{Oliveira}.
\newblock {Physical meaning and measurement of the entropic parameter q in an
  inhomogeneous magnetic systems}.
\newblock {\em European Physical Journal B}, 50:99, 2006.

\bibitem{1999_BrJP_29_1}
C.~{Tsallis}.
\newblock {Nonextensive Statistics: Theoretical, Experimental and Computational
  Evidences and Connections}.
\newblock {\em Brazilian Journal of Physics}, 29:1, 1999.

\bibitem{2004_CMT_16_223}
C.~{Tsallis} and E.~{Brigatti}.
\newblock {Nonextensive statistical mechanics: A brief introduction}.
\newblock {\em Continuum Mechanics and Thermodynamics}, 16:223, 2004.

\bibitem{livro_TsallisCap1}
C.~Tsallis.
\newblock {\em Nonextensive Entropy - Interdisciplinary Applications}, chapter
  Nonextensive statistical mechanics: Construction and physical interpretation.
\newblock eds. M.~Gell-Mann and C.~Tsallis. Oxford University Press, New York,
  2004.

\bibitem{2005_PNAS_102_15377}
C.~{Tsallis}, M.~{Gell-Mann}, and Y.~{Sato}.
\newblock {Asymptotically scale-invariant occupancy of phase space makes the
  entropy Sq extensive}.
\newblock {\em Proceedings of the National Academy of Science}, 102:15377,
  2005.

\bibitem{website_TEMUCO}
For a complete and updated list of references, see the web site:
  \texttt{tsallis.cat.cbpf.br/biblio.htm}.

\bibitem{1999_PhysA_268_553}
A.~R. {Lima}, J.~S.~S. {Martins}, and T.~J.~P. {Penna}.
\newblock {Monte Carlo Simulation of Magnetic System in the Tsallis
  Statistics}.
\newblock {\em Physica A}, 268:553, 1999.

\bibitem{1991_JPA_24_69}
E.~M.~F. {Curado} and C.~{Tsallis}.
\newblock {Generalized statistical mechanics: connection with thermodynamics}.
\newblock {\em Journal of Physics A}, 24:L69, 1991.
\newblock Corrigenda: 24 (1991) 3187 and 25 (1992) 1019.

\bibitem{1998_PhysA_102_15377}
C.~{Tsallis}, R.~S. {Mendes}, and A.~R. {Plastino}.
\newblock {The role of constraints within generalized nonextensive statistics}.
\newblock {\em Physica A}, 261:534, 1998.

\bibitem{livro_beck}
C.~Beck and F.~Schlogl.
\newblock {\em Thermodynamics of Chaotic Systems: An Introduction}.
\newblock Cambridge University Press, Cambridge, 1993.

\bibitem{2005_JSMTE_04_009}
G.~L. {Ferri}, S.~{Mart{\'{\i}}nez}, and A.~{Plastino}.
\newblock {Equivalence of the four versions of Tsallis's statistics}.
\newblock {\em Journal of Statistical Mechanics: Theory and Experiment},
  4:P04009, 2005.

\bibitem{2000_PhysA_283_59}
R.~{Salazar} and R.~{Toral}.
\newblock {A Monte Carlo Method for the Numerical Simulation of Tsallis
  Statistics}.
\newblock {\em Physica A}, 283:59, 2000.

\bibitem{2001_PhysA_295_246}
S.~{Mart{\'{\i}}nez}, F.~{Pennini}, and A.~{Plastino}.
\newblock {The concept of temperature in a nonextensive scenario}.
\newblock {\em Physica A}, 295:246, 2001.

\bibitem{2001_PhysA_295_416}
S.~{Mart{\'{\i}}nez}, F.~{Pennini}, and A.~{Plastino}.
\newblock {Thermodynamics' zeroth law in a nonextensive scenario}.
\newblock {\em Physica A}, 295:416, 2001.

\bibitem{2002_PhysA_305_52}
R.~{Toral} and R.~{Salazar}.
\newblock {Ensemble equivalence for non-extensive thermostatistics}.
\newblock {\em Physica A}, 305:52, 2002.

\bibitem{2003_PhysA_317_209}
R.~{Toral}.
\newblock {On the definition of physical temperature and pressure for
  nonextensive thermostatistics}.
\newblock {\em Physica A}, 317:209, 2003.

\bibitem{2004_EPL_65_606}
Q.~A. {Wang}, L.~{Nivanen}, A.~{LeM{\'e}haut{\'e}}, and
M.~{Pezeril}.
\newblock {Temperature and pressure in nonextensive thermostatistics}.
\newblock {\em Europhysics Letters}, 65:606, 2004.

\bibitem{2006_PhysA_368_430}
S.~{Abe}.
\newblock {Temperature of nonextensive systems: Tsallis entropy as Clausius
  entropy}.
\newblock {\em Physica A}, 368:430, 2006.

\bibitem{2002_EPL_58_42}
M.~S. {Reis}, J.~C.~C. {Freitas}, M.~T.~D. {Orlando}, E.~K.
{Lenzi}, and I.~S.
  {Oliveira}.
\newblock {Evidences for Tsallis non-extensivity on CMR manganites}.
\newblock {\em Europhysics Letters}, 58:42, 2002.

\bibitem{2002_PRB_66_134417}
M.~S. {Reis}, J.~P. {Ara{\'u}jo}, V.~S. {Amaral}, E.~K. {Lenzi},
and I.~S.
  {Oliveira}.
\newblock {Magnetic behavior of a nonextensive S-spin system: Possible
  connections to manganites}.
\newblock {\em Physical Review B}, 66:134417, 2002.

\bibitem{2003_PRB_68_014404}
M.~S. {Reis}, V.~S. {Amaral}, J.~P. {Ara{\'u}jo}, and I.~S.
{Oliveira}.
\newblock {Magnetic phase diagram for a nonextensive system: Experimental
  connection with manganites}.
\newblock {\em Physical Review B}, 68:014404, 2003.

\bibitem{livro_compphys}
R.~H. Landau and M.~J. Paez.
\newblock {\em Computational Physics: Problem solving with computers}.
\newblock Wiley-VHC, Weinheim, 2004.

\bibitem{2000_PhysA_286_156}
I.~{Bediaga}, E.~M.~F. {Curado}, and J.~M. {de Miranda}.
\newblock {A nonextensive thermodynamical equilibrium approach in
  ${e^{+}e^{-}\rightarrow}$ {\it hadrons}}.
\newblock {\em Physica A}, 286:156, 2000.

\bibitem{1999_PRL_83_4233}
R.~{Salazar} and R.~{Toral}.
\newblock {Scaling Laws for a System with Long-Range Interactions within
  Tsallis Statistics}.
\newblock {\em Physical Review Letters}, 83:4233, 1999.

\bibitem{2000_EPJB_17_679}
R.~{Salazar}, A.~R. {Plastino}, and R.~{Toral}.
\newblock {Weakly nonextensive thermostatistics and the Ising model with
  long-range interactions}.
\newblock {\em European Physical Journal B}, 17:679, 2000.

\bibitem{2001_PhysA_290_159}
R.~{Salazar} and R.~{Toral}.
\newblock {Thermostatistics of extensive and non-extensive systems using
  generalized entropies}.
\newblock {\em Physica A}, 290:159, 2001.

\end{thebibliography}
\end{document}